\documentclass[twocolumn,showpacs,showkeys,preprintnumbers,amsmath,amssymb]{revtex4}

\usepackage{graphicx}
\usepackage{dcolumn}
\usepackage{bm}
\usepackage{epsfig}
\usepackage{psfrag}
\usepackage{upgreek}

\begin{document}

\preprint{Final draft}

\title{The Bouncing Jet:\\A Newtonian Liquid Rebounding off a Free Surface}
\author{Matthew Thrasher}
 \email{thrasher@chaos.utexas.edu}
\author{Sunghwan Jung}
\altaffiliation[Current address:]{ Mathematics Department, MIT}
\author{Yee Kwong Pang}
\altaffiliation[Current address:]{ Hong Kong University of Science
and Technology.}
\author{Chih-Piao Chuu}
\author{Harry L. Swinney}
 \email{swinney@chaos.utexas.edu}
\affiliation{%
Center for Nonlinear Dynamics and Department of Physics, University
of Texas at Austin, Austin TX 78712}

\date{\today}

\begin{abstract}
We find that a liquid jet can bounce off a bath of the same liquid
if the bath is moving horizontally with respect to the jet. Previous
observations of jets rebounding off a bath (e.g. Kaye effect) have
been reported only for non-Newtonian fluids, while we observe
bouncing jets in a variety of Newtonian fluids, including mineral
oil poured by hand. A thin layer of air separates the bouncing jet
from the bath, and the relative motion replenishes the film of air.
Jets with one or two bounces are stable for a range of viscosity,
jet flow rate and velocity, and bath velocity.  The bouncing
phenomenon exhibits hysteresis and multiple steady states.

\end{abstract}



\maketitle
\section{\label{sec:Introduction}Introduction}

\begin{figure}[htbp]
\begin{center}
  \includegraphics[width=8.7cm]{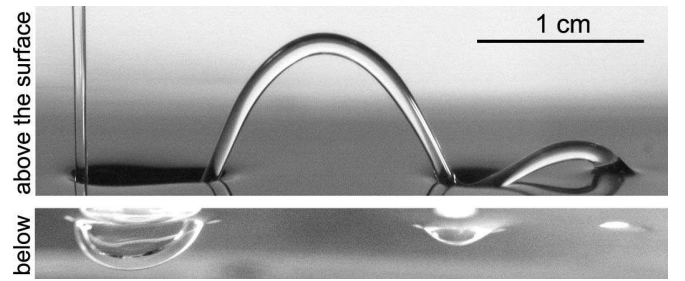}
\end{center}
\caption{A liquid jet bounces twice before merging with the bath,
which is moving to the right. The jet and bath are silicone oil of
the same viscosity.  The upper and lower pictures were taken from
above and from below the bath surface; the images were not obtained
at the same time or the same angle, so small differences exist. The
jet's image can be seen reflected on the surface. Parameters: liquid
viscosity $\mu$ = 102 mPa s (about 100 times more viscous than
water), jet flow rate $Q$ = 0.35 cm$^3$/s, falling height $H$ = 5.0
cm, and horizontal velocity of the bath $V_{bath}$ = 15.7 cm/s.}
 \label{DoubleBounce}
\end{figure}

A liquid stream falling onto the free surface of a liquid bath can
merge immediately on contact, plunge through the surface and entrain
air \cite{entrap:eggers,entrap:quere}, coil up like a rope
\cite{rope:bonn},  float on the surface prior to coalescing
\cite{noncoal:couder}, float without ever coalescing
\cite{noncoal:review}, or break into droplets \cite{b:breakup2}. We
report in this paper observations of a jet of Newtonian liquid
bouncing off a horizontally moving surface of a bath of the same
liquid. Figure \ref{DoubleBounce} shows a typical bouncing jet
viewed from the side.  The jet falls to the bath's surface, is bent
upwards, and undergoes a short flight. After rebounding once more
off the surface, the stream merges with the bath.  In all figures
the liquid bath is moving to the right, and the stream and bath are
the same fluid.  This paper examines when and how a liquid jet
bounces. The issues that arise in studying the bouncing jet (e.g.
non-coalescence, lubrication, and entrainment) are ubiquitous in
fluid processing, such as pouring and mold casting. They also are
critical in the design of bearings \cite{noncoal:suppress},
gas-liquid reactors \cite{entrap:review}, film coating equipment
\cite{entrap:coatings}, and metallurgical procedures
\cite{b:breakup2}.

Drops of liquid floating and bouncing on the surface of a bath have
been studied scientifically for over 125 years
\cite{noncoal:rayleigh1,noncoal:dropss,noncoal:drops,drops:bouncing}.
On a pond during a light rainfall, splashing  raindrops throw up
smaller drops, and these smaller drops often can be seen to sit on
the water surface momentarily. During this time of noncoalescence, a
thin layer of air separates each drop from the pond. Noncoalescence
can be prolonged (sometimes indefinitely) by either replenishing the
air between the two liquid bodies or slowing the loss of the
existing air. This can be achieved with surfactants
\cite{noncoal:dropssurfactant,drops:dancing}, vibration
\cite{noncoal:couder}, microgravity \cite{noncoal:microgravity}, a
velocity difference between the drop and bath
\cite{noncoal:suppress}, evaporation \cite{noncoal:review2},
thermo-capillarity \cite{noncoal:review2}, or by increasing the
viscosity of the surrounding medium \cite{noncoal:mahajan2}.

The bouncing of a liquid jet has also been observed for a fluid of
elongating polymers incident at a glancing angle on a rotating
drum~\cite{kaye:nonnewton}. Bouncing also occurs for a jet of
shear-thinning liquid falling onto a pool of the same liquid; this
is called the Kaye effect \cite{kaye:orig}.  While this effect is
visually similar to the bouncing jet phenomenon presented in this
paper, the Kaye effect occurs only in non-Newtonian fluids
\cite{kaye:lohse} (see discussion in Section \ref{sec:Discussion}).

Jets of water colliding midair at a glancing angle can also bounce
off each other, because during the collision they are separated by a
layer of air ~\cite{noncoal:rayleigh3,noncoal:review}. However, no
systematic study has been conducted of Newtonian jets bouncing off a
bath surface.  To measure the conditions necessary for the bouncing
of a Newtonian liquid jet, we built an experimental apparatus
(Section \ref{sec:Experiment}), and we found that jets bounce for a
wide range of parameters (Section \ref{sec:Results}).  By comparing
the energies associated with the non-bouncing and bouncing states,
we suggest why bouncing is preferred rather than plunging (Section
\ref{sec:Scaling}). A bouncing jet can easily be reproduced with
common materials (Section \ref{sec:Kitchen}). These observations are
related to previous work in Section \ref{sec:Discussion}.

\begin{figure}[htbp]
\begin{center}
  \includegraphics[width=8cm]{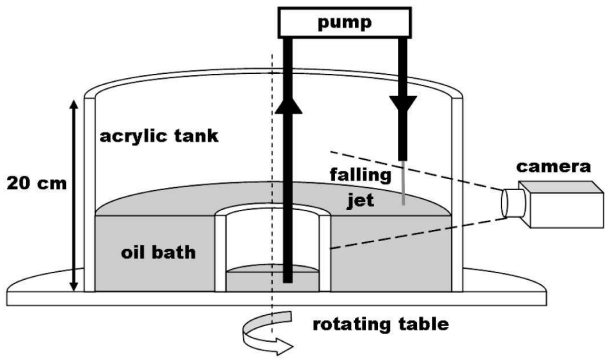}
\end{center}
\caption{Experimental setup: a bath of silicone oil is rotated under
a falling stream of the same oil. A camera in the laboratory frame
records the motion through the tank's clear acrylic side.}
\label{Apparatus}
\end{figure}

\begin{figure*} 
\begin{center}
  \includegraphics[width=\linewidth]{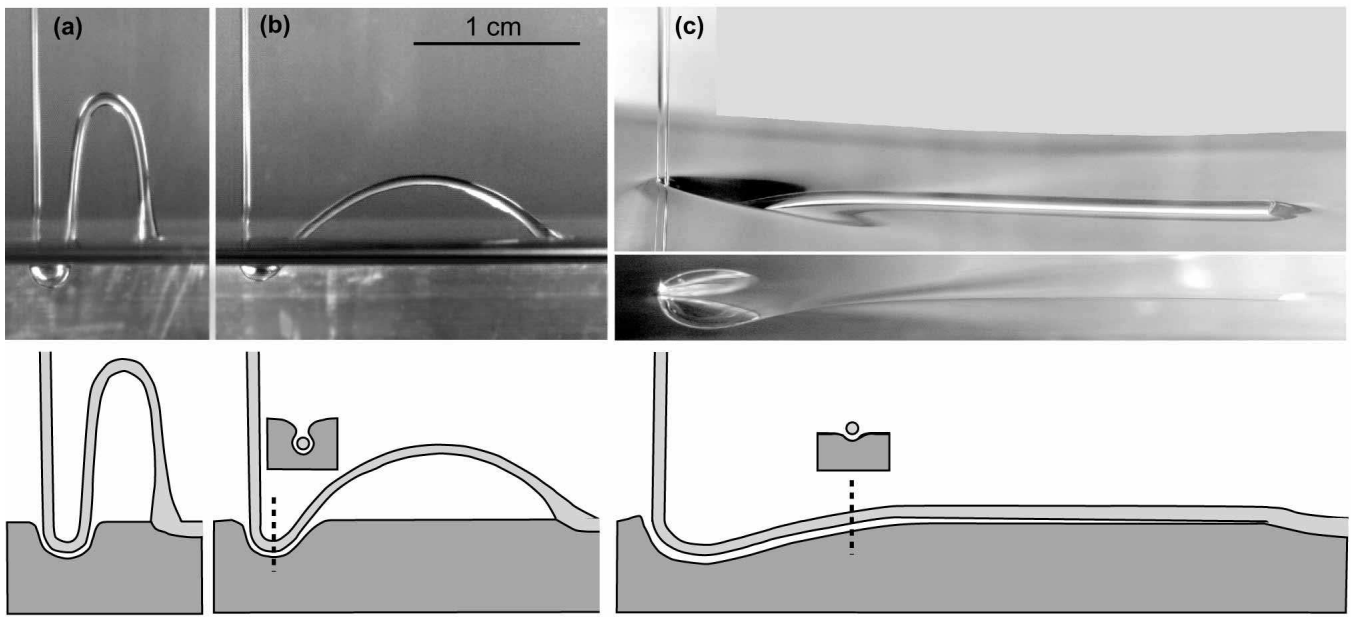}
\end{center}
\caption{ Photographs (top) and schematic drawings (bottom) of the
jet for different horizontal velocities of the bath, $V_{bath}$: (a)
0.656 cm/s, a jet bouncing nearly vertically. (b) 3.28 cm/s, a jet
bouncing more obliquely. (c) 42.3 cm/s, a trailing (non-bouncing)
jet. Increasing $V_{bath}$  imparts more horizontal momentum to the
rebounding jet.  The schematics exaggerate the thickness of the
layer of air between the jet and bath. For (a) and (b), the dark
horizontal line is the surface's meniscus on the outer tank wall; in
(c), the top and bottom images were taken above and below the
surface, so that the surface extends back in perspective. Parameters
in (a) and (b) were $\mu$ = 349 mPa s, $Q$ = 0.16 cm$^3$/s, $H$ =
4.2 cm; in (c), $\mu$ = 102 mPa s, $Q$ = 0.35 cm$^3$/s, $H$ = 5.0
cm.}
 \label{ChangingV}
\end{figure*}

\section{\label{sec:Experiment}Experiment}

We used a rotating annulus of fluid (Fig.~\ref{Apparatus}) to
maintain a constant horizontal velocity of a bath with respect to a
vertically impinging jet. The parameters varied were the viscosity
$\mu$ of the silicone oils, the jet's flow rate $Q$, the height of
the nozzle $H$ above the bath surface, and the relative velocity
$V_{bath}$ of the bath to the nozzle.

A cylindrical tank with a clear acrylic outer wall was mounted on a
rotating table.  The annular bath was 39.1 cm in outer diameter,
27.3 cm in inner diameter, and 7.7 cm deep. Silicone oils were used
for their stability, low surface tension, high viscosity, and
Newtonian properties.  They were Dow Corning 200$\textregistered$
series and Clearco oils with viscosity $\mu =$ 52 to 349  mPa s,
density $\rho =$ 959 to 968 kg/m$^3$, and surface tension $\sigma =$
21.0 to 21.2 mN/m.  Measurements were made at the bath temperature
of 23 $\pm$ 1$^\circ$ C.  We measured the viscosity of each oil for
shear rates from 1 to 10$^4$ s$^{-1}$ with a Paar Physica MCR300
rheometer, and we found at most a weak dependence on shear rate:
oils with viscosities of $\mu$ = 52, 102, 211, and 349 mPa s at a
low shear rate had viscosity values 2, 4, 8, and 12\% lower at
10$^4$ s$^{-1}$, respectively.

The typical shear rate in the liquid in our experiments is difficult
to estimate because the velocity profiles in the air and liquid were
not measured.  Most of the shear was in the air layer since the
dynamic viscosity ratio of air to oil ranged from 4 $\times$
10$^{-4}$ to 5 $\times$ 10$^{-5}$.  Even ignoring the air layer, the
largest velocity difference [1.7 m/s, see
Fig.~\ref{fig:PhaseDiagrams}(b)] across the smallest jet diameter
(0.05 cm) would produce a maximum shear rate of 3400 s$^{-1}$, at
which even the most viscous oil decreased in viscosity by only a few
percent.  The actual shear rates in the liquid phase should be much
smaller; hence the silicone oils used can be considered as Newtonian
fluids for the conditions in the experiment.

The table's rotation rate determined the relative velocity between
the bath surface and the nozzle ($V_{bath} \equiv \Omega R$ ranged
from less than 1 cm/s to 35 cm/s with a typical distance from the
rotation axis $R =$ 16 cm and a rotation rate $\Omega/2\pi =$ 0 to
0.4 Hz).   When changing the rotation rate, adequate time was given
for the bath to establish solid-body rotation.  The typical
uncertainty in the horizontal bath velocity was 1\%.

The flow rate $Q$ was controlled by a gear pump with a pulse
dampener and a bypass; $Q$ ranged from 0.075 to 6.28 cm$^3$/s with a
typical uncertainty of 2\%. Excess liquid drained over an interior
wall into a central reservoir from which liquid was pumped; thus the
bath's surface height was constant at the inner cylinder, barring
interfacial pinning.  The rotation rate changed the bath level at
the jet's position by at most 0.1 cm.

The pump withdrew oil from the central basin and released it above
the surface through a vertical Teflon nozzle (stationary in the
laboratory frame). The nozzle had an inner diameter $d_{nozzle}$ =
0.52 cm and produced a vertical liquid stream at a height $H$ above
the liquid surface; $H$ ranged from 0.7 cm to 15 cm, as determined
(within 3\%) using a cathetometer. The velocity of the jet
$V_{jet}$, measured at the point of first impact with the surface,
was mostly due to falling from height $H$ and only changed slowly
with $Q$ (i.e. $4 Q/\pi d_{nozzle}^2 << \sqrt{2 g H}$). With typical
values $V_{jet}$ = 60 cm/s and the jet diameter $d_{jet}$ = 0.1 cm,
the Reynolds number of the jet was 6, the Bond number was 0.5, the
capillary number was 3, and the Weber number was 18.  Dust and
bubbles sitting on the surface could destabilize a bouncing jet, so
they were removed by dragging a mesh over the surface of the bath
opposite to the falling jet (not shown in Fig.~\ref{Apparatus}).

The bouncing was initiated by passing a small (0.6 cm diameter)
horizontal rod quickly through the falling jet, changing its radius,
velocity, and shape of the jet in a complicated, time-dependent
manner.  When the non-uniformity collided with the bath surface, a
non-bouncing jet often started to bounce.

Two other methods were found to initiate the bouncing, but were not
used in mapping the regime diagrams in Section
\ref{subsec:phasediagrams}.  One method was to rapidly decrease the
flow rate from a high rate that entrained air.  As the flow rate
decreased, the submerged jet penetrated less deeply and then began
to bounce.  The other method was to change the bath velocity.  This
last method is discussed more in Section \ref{subsec:multiplicity}.

Images were acquired in the lab frame through the outer wall of the
tank with a digital camera. If an ambiguity existed in the geometry
of the jet and bath, images were taken at several angles. The images
were used to measure the diameter of the jet, and the velocity was
computed by using the flow rate and continuity. The typical
uncertainty of the vertical velocity measurement was 8\%.

\begin{figure}[htbp]
\begin{center}
  \includegraphics[width=8.7cm]{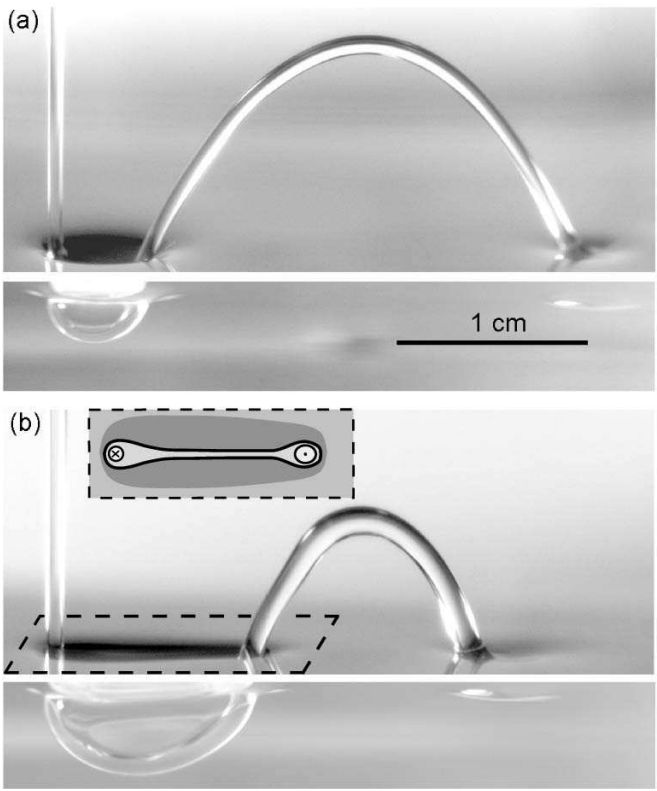}
\end{center}
\caption{Photographs of the bouncing jet for flow rates $Q$: (a)
0.23 cm$^3$/s, (b) 0.44 cm$^3$/s; the top and bottom pictures in
each case are respectively above and below the surface. The jet with
the greater flow rate pushes the surface deeper and the jet rebounds
lower. Each jet rebounds moving slower than its initial speed (so
that by continuity, the jet becomes thicker).  The inset in (b) is a
top view of the area with the dashed outline; this region in the
picture is just below the bath level and is in shadow. The closed
black curves within the inset show a horizontal cross section of the
interfaces just below the bath level. Parameters for both (a) and
(b): $\mu$ = 102 mPa s, $H$ = 5.0 cm, $V_{bath}$ = 15.7 cm/s.}
\label{fig:PumpRateChange}
\end{figure}

\section{\label{sec:Results}Results}

\subsection{\label{subsec:VbandQ}Dependence on bath velocity and flow rate}

Two important parameters that change the qualitative behavior of the
jet are the bath velocity and the flow rate. At low bath velocity,
the jet bounced with a nearly vertical rebound, as in
Fig.~\ref{ChangingV}(a). At smaller bath velocity, the jet bounced
only intermittently because the bouncing liquid would collide with
the falling jet or would distort the surface and destabilize the
bouncing.  To get a stable bounce the bath velocity had to be fast
enough to carry away the rebounding fluid so it would not disturb
the impinging jet. The bouncing could stop in another way: the
height of the jet's bounce would quickly decrease until the jet
merged with the bath.

As the bath velocity increased, the rebound became more oblique, as
in Fig.~\ref{ChangingV}(b). As a function of the bath's horizontal
velocity, the jet gained more horizontal momentum from the bath and
the angle of the jet's rebound changed continuously.

During bouncing, the jet and the bath were separated by a
lubricating layer of air, as revealed by a laser beam propagating
down the falling jet.  The beam was internally reflected within the
jet while the jet bounced, and entered the bath only when the jet
and bath merged afterward. We did not measure the thickness of the
air layer, but for a plunging jet the air layer surrounding a jet
has been measured by Lorenceau \emph{et al.}~\cite{entrap:quere} to
be several micrometers thick. For a rebounding water drop, the
minimum thickness of the air film was calculated by Jayaratne and
Mason \cite{noncoal:mason} to be about 0.1 $\upmu$m.  For a pendant
drop suspended above a moving solid surface, the thickness of the
film was measured by Vetrano and Dell'Aversana to be a few
micrometers on average, and the shape of the film did not change
significantly if the ambient pressure of the surrounding air was at
least 300 to 400 mbar \cite{noncoal:review2}.

The velocity of the bath necessary for a stable bounce was small
compared to the jet velocity. Defining the jet incidence angle as
$\theta = \textnormal{tan}^{-1}(V_{jet}/V_{bath})$, we found that
$\theta$ only ranged from 83 to 90$^\circ$, while the angle of
rebound (in the bath's frame) ranged from 20 to 80$^\circ$ (for
$\mu$ = 349 mPa s, $Q$ from 0.16 to 0.52 cm$^3$/s, $H$ = 4.2 cm, and
$V_{bath}$ from 0.7 to 7.9 cm/s). For these conditions, 13 to 29\%
of the jet's speed was lost while bouncing. Typically, the jet
rebounded with higher speed with increasing $V_{bath}$ until the jet
rebounded low enough for its increasing contact with the bath to
slow the jet substantially.  This trend is consistent with
measurements made by Jayaratne and Mason of the rebound of
individual water drops from a water surface \cite{noncoal:mason}.

At a higher bath velocity, the jet no longer lifted off the surface
of the bath, but rather floated on top of the bath, as in
Fig.~\ref{ChangingV}(c).  We call this state the ``trailing jet.''
The trailing jet could be readily identified by looking from below
the surface at the continuous indentation made by the jet on the
surface.   The jet did not coalesce with the bath until the thin
layer of air between it and the bath drained enough to burst or was
disturbed, such as by dust or an irregularity in the jet. The
trailing jet often collapsed in long sections at a time, which
indicated that the air film had ruptured in the beginning or middle
of the trailing jet.

The transition between a bouncing jet and a trailing jet was abrupt
in some cases and gradual in others.  For example, at $\mu$ = 102
mPa s and $Q$ between 0.10 and 1.05 cm$^3$/s, the jet would no
longer bounce above a particular bath velocity. Instead, the jet
would trail along the surface; the length of liquid floating on the
surface quickly shortened until the jet merged with the bath. On the
other hand, for $\mu$ = 211 mPa s and $Q$ between 0.55 and 0.94
cm$^3$/s, the jet would lift off the surface less and less as
$V_{bath}$ increased, until a jet of constant length floated on the
surface.

The bouncing jet's behavior also depends on flow rate, as
Fig.~\ref{fig:PumpRateChange} illustrates. The higher the flow rate,
the more vertical momentum the jet has to deform the bath's surface.
This leads to a deeper, ellipsoidal indentation and more viscous
drag on the jet by the bath. The decrease in jet velocity can be
seen by the thickening of the jet and the smaller bounce height. The
bouncing jet in Fig.~\ref{fig:PumpRateChange}(b) is slightly
irregular. Irregularities in the pumping, nozzle position, and
surrounding air can cause the bouncing jet to be temporarily
unsteady. The unsteady motion usually decays back to the steady
bounding jet, but was sometimes observed to be sustained and
periodic.


During rebound, the jet is below the bath's surface level for some
distance; the jet's weight and the changing momentum of the jet are
balanced by surface tension and buoyancy [see the inset of the
schematic in Fig.~\ref{ChangingV}(b)].  However, for most jets, the
buoyancy of the indentation is small because the sides of the
surface indentation are close together and little volume is
displaced [see the inset of Fig.~\ref{fig:PumpRateChange}(b)].  Also
the jet's weight can be neglected, because of the large change in
jet's momentum.  In this case the force changing the jet's momentum
is provided mainly by surface tension. The surface pulls nearly
vertically on the length $\ell$ of the jet that is under the bath's
surface, producing a force $F_S \approx 2 \sigma \ell$. Assuming
that the jet velocity is the same before and after rebound, the rate
of change of the jet's vertical momentum is $F_{I} = \frac{\Delta
p_y}{\Delta t_c} \approx \rho \pi (d_{jet}/2)^2 V_{jet}^2 (1 + sin
\phi)$, where $\Delta p_y$ is the change in the jet's momentum in
the vertical direction, $\Delta t_c$ is the duration of the
collision, and $\phi$ is the angle of rebound measured from the
horizon. In Fig.~\ref{fig:PumpRateChange}(b), the length $\ell$ =
1.3 $\pm$ 0.1 cm and the angle $\phi$ = 70 $\pm$ 2$^{\circ}$.
Therefore, for this simplified force calculation, the surface force
$F_S$ = 54 $\pm$ 4 dynes and the force of the jet changing direction
$F_I$ = 52 $\pm$ 6 dynes, where only the measurement uncertainties
are included here. The systematic errors from the approximations
made in this argument are not included.  With the approximations,
the forces might be expected to agree within a factor of 2.  The
forces are the equal within the measurement uncertainty, perhaps
coincidentally.

\subsection{\label{subsec:phasediagrams}Regime diagrams}
\begin{figure}
\begin{center}
\includegraphics[width=9.2cm]{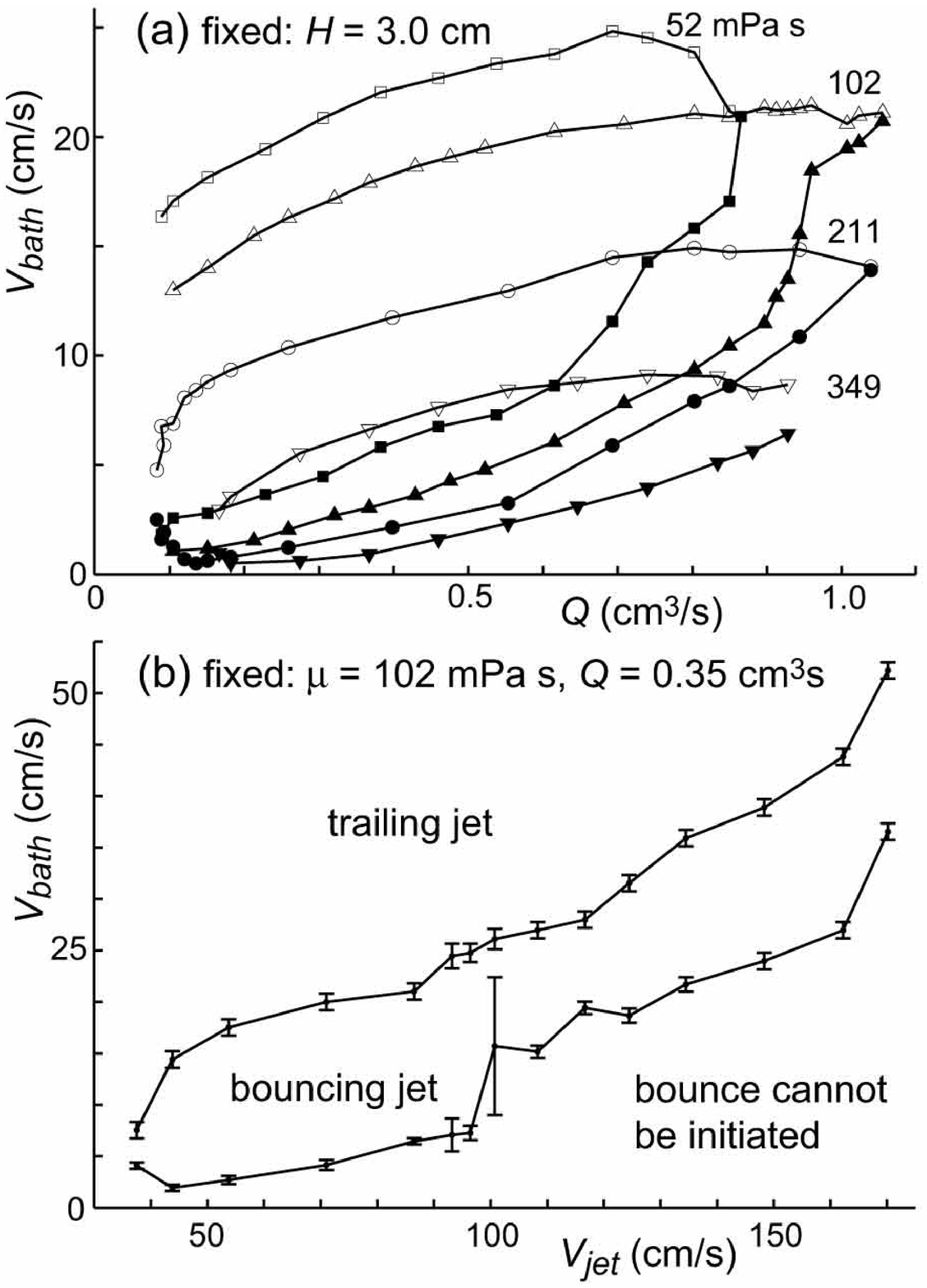}
\end{center}
\caption{Conditions for which the jet bounces: (a) Bounces could be
initiated between the lines with closed and open points but could
not be initiated below the lines with closed points or above the
open points; a trailing jet was often observed above the open points
(viscosity values $\mu$ = 53 ($\blacksquare,\square$), 102
($\blacktriangle,\vartriangle$), 211 ($\bullet,\circ$), and 349 mPa
s ($\blacktriangledown,\triangledown$).)  The bath velocity at which
the transitions occur were reproducible typically within 4\%; the
uncertainty was greater where the transition curves have a steeper
slope. (b) Dependence of the region of stable bounce on $V_{jet}$
and $V_{bath}$; the range of $V_{jet}$ (38 to 170 cm/s) corresponds
to $H$ values 1.7 to 14.1 cm.
Error bars show the uncertainty of each transition point. Around
$V_{jet} \approx$ 100 cm/s, there is a greater uncertainty
corresponding to a transition from a non-entraining jet to an
air-entraining jet as $V_{jet}$ is increased.}
\label{fig:PhaseDiagrams}
\end{figure}

Sweeps of the parameters $\mu, Q, H, V_{bath}$ were conducted with
either $H$ held constant [Fig.~\ref{fig:PhaseDiagrams}(a)] or $\mu$
and $Q$ held constant [Fig.~\ref{fig:PhaseDiagrams}(b)]. Each point
was measured at least three times. Bouncing was initiated by passing
a plastic rod through the falling liquid stream. A bouncing stream
was considered ``initiated'' after 5 s; most jets that were stable
for 5 s would persist for much longer times. The initiation
procedure yielded transition parameter values that were reproducible
and consistent for different experimenters. The choice of 5 s
persistence for identifying a stable bounce is arbitrary; a
criterion of 1 s duration would yield parameter space regions for
bouncing somewhat larger than those in Fig.~\ref{fig:PhaseDiagrams}.

Figure \ref{fig:PhaseDiagrams}(a) displays the range of $V_{bath}$
for which a bounce could be initiated as a function of $Q$ (for a
fixed nozzle height, $H = 3.0$ cm). For each oil viscosity value,
there is a transition between no bouncing and bouncing, marked by
solid points in Fig.~\ref{fig:PhaseDiagrams}(a). The open points
mark the greatest bath velocity where bouncing could be initiated.
Above the open points, the jet often trailed on the surface either
temporarily or steadily.

The regime where bouncing could be initiated did not close at low
$Q$ for the viscosities of $\mu =$ 52 and 102 mPa s; the jet broke
into droplets before reaching the surface.  To prevent dripping,
care was taken that the oil did not wet the Teflon nozzle beyond the
rim of the its opening; however, dripping could not be prevented for
low $Q$.  For $\mu=$ 349 mPa s at high $Q$, the bouncing regime did
not close because the jet no longer lifted off the surface for
higher flow rates. At viscosities higher than 349 mPa s, the
transitions from non-bouncing to bouncing became difficult to
reproduce because of sensitivity to mechanical vibration. At some
conditions, mechanical noise kept the jet bouncing, while at other
conditions, the same amount of noise destabilized the bouncing jet.

Jets can bounce twice, as in Fig.~\ref{DoubleBounce}, but this
occurs in a smaller parameter space region than the jet undergoing a
single bounce (Fig.~\ref{fig:PhaseDiagrams}); the region of double
bouncing was not mapped.

The region of stable bouncing when $H$ rather than $Q$ was varied is
shown in Fig. \ref{fig:PhaseDiagrams}(b). Roughly, for higher $H$
(and hence higher $V_{jet}$), a higher horizontal bath velocity is
needed for stable bouncing.

The range in which stable bouncing occurs is hysteretic in two
senses. First, the region in which the jet bounces is larger if the
experimental parameters are changed after a bounce has been
initiated. For example, if the bath's velocity was decreased slowly
while a jet was bouncing, the bath velocity at which the jet stopped
bouncing was lower than the bath velocity at which the bouncing jet
could be initiated. The second sense of hysteresis is that a jet
impinging on a moving bath has up to four distinct states that occur
for the same experimental conditions.

\subsection{Multiple stable states\label{subsec:multiplicity}}

\begin{figure*}
\begin{center}
\includegraphics[width=\linewidth]{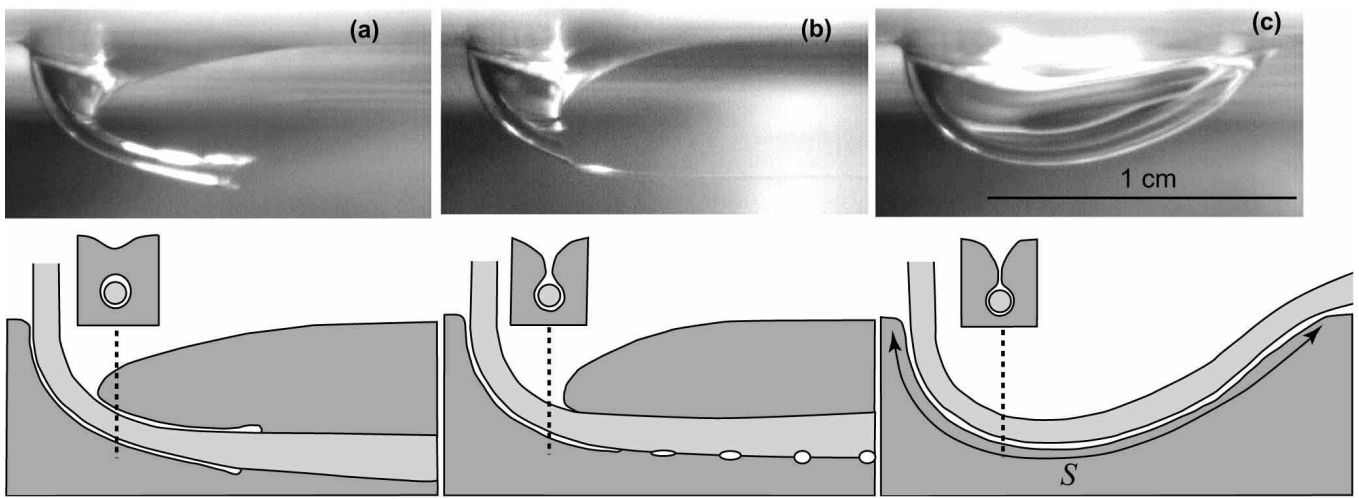}
\end{center}
\caption{Multiplicity: the three states (a)-(c) are stable for the
same conditions. The bath surface is at the top of each photograph,
and below each photograph is a corresponding schematic diagram. (a)
Plunging jet with air entrainment and no bounce; bubbles break off
from the top edge of the entrained air sheath. (b) Half-entraining
jet, which entrains air only on its lower edge; bubbles (blurred to
a faint line by the long exposure) break off from the entrained air
skirt. (c) Bouncing jet, whose curved surface refracts light and
produces a distorted image of the background. Parameters: $\mu$ =
106 mPa$\cdot$s, $Q$ = 0.35 cm$^3$/s,  $H$ = 6.4 cm, $V_{bath}$ =
15.2 cm/s.} \label{fig:Multiplicity}
\end{figure*}

Three states that are stable for the same conditions are shown in
Fig.~\ref{fig:Multiplicity}. Each state can be changed to another
state by passing a plastic rod through the falling stream.

Figure \ref{fig:Multiplicity}(a) shows a thin cylindrical film of
air being entrained into the bath by the impinging jet.  (An
impinging jet in a \emph{stationary} bath was studied by Lorenceau
\emph{et al.} \cite{entrap:quere,entrap:fracture}.) The horizontal
motion of the bath drags along the jet and the air film. Air is
entrained continuously and collects at the end of the sheath;
occasionally a bubble pinches off.

A second state, which we call the ``half-entraining jet'', is shown
in Fig.~\ref{fig:Multiplicity}(b). Only the bottom edge of the jet
entrains air; the top edge of the jet is deep in the bath but does
not entrain air, as indicated by the inset of the schematic diagram.
As the entrained air collects, small bubbles separate from the
bottom edge of the air sheath.

Another confirmation that an air layer separates the jet and the
bath is given by increases in the bath velocity, which lengthen the
air sheath of the half-entraining jet into a nearly semi-circular
arc until the jet rises above the bath level and floats on the
surface. Since the rotation rate could be adjusted continuously,
this transition could be approached slowly.  Once the jet is
trailing on the surface, decreasing the bath velocity can lead to
the jet lifting from the bath surface; this was mentioned previously
as the third method to start a jet bouncing.

The bouncing jet in Fig.~\ref{fig:Multiplicity}(c) is the third
state for the same conditions. The jet deforms the bath's surface
for a distance $S$. The geometry is difficult to deduce from the
photograph because of the refraction, but it is illustrated in the
inset of the schematic diagram. As $V_{bath}$ increased, the jet
plunged less deeply below the free surface.  The maximum depth that
the jet penetrated into the bath decreased linearly as the bath's
velocity $V_{bath}$ was increased from 0.7 to 9.43 cm/s, while the
horizontal distance that the jet traveled below the bath level
increased (observations for $Q$ = 0.16 to 0.52 cm$^3$/s; $\mu$ = 349
mPa s, $H$ = 4.2 cm).   As $V_{bath}$ increased, the decrease in the
penetration depth was greater than the increase in the horizontal
distance, so that the length scale $S$ of the bouncing jet decreased
slowly.

In a fourth state, observed for some conditions but not those in
Fig.~\ref{fig:Multiplicity}, the jet merged with the bath upon
contact with the bath, causing only a small depression of the bath
surface around the circumference of the jet. The conditions for
merging smoothly and for entraining air were discussed in
\cite{entrap:eggers}.

\section{\label{sec:Scaling}Energy of plunging and bouncing jets}

We suggest that the transition from plunging to bouncing,
illustrated in Fig.~\ref{fig:Multiplicity} and marked by the lower
curves in Fig.~\ref{fig:PhaseDiagrams}(a), corresponds to a
competition between the work associated with dragging the plunging
jet through the bath and the energy needed to create the additional
bath surface when the jet is bouncing. Effects such as buoyancy,
inertia, and shear in the air film are important in the initiation
and process of bouncing but are assumed to be unimportant for the
argument.

\begin{figure}[htbp]
\begin{center}
  \includegraphics[width=\linewidth]{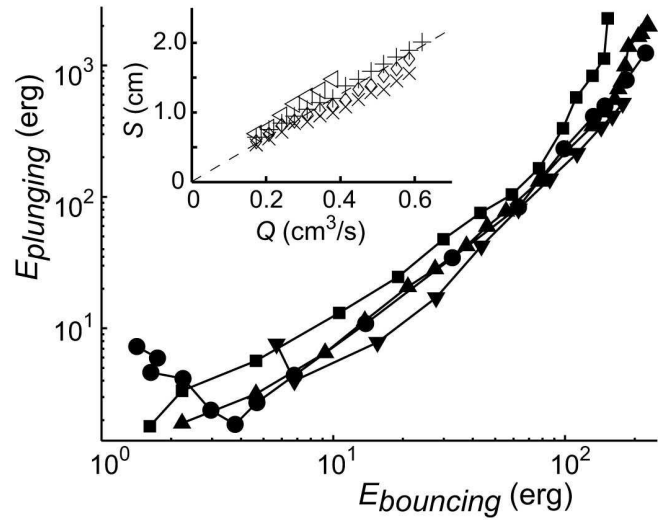}
\end{center}
\caption{It is suggested that plunging is energetically favored
below the curves while bouncing is energetically favored above the
curves. The argument is supported by the collapse of the data for
four viscosity values (ranging by a factor of seven) onto the same
approximate curve.  The energy $E_{plunging}$, given in
(\ref{E_plunging}), is from the viscous drag of the plunging jet.
The energy $E_{bouncing}$, given in (\ref{E_bouncing}), is from the
surface area of the bouncing jet. The solid symbols and data points
are the same as in Fig.~\ref{fig:PhaseDiagrams}(a).  Inset: the
penetration length $S$ varies linearly with $Q$ ($\mu = 211$ mPa s)
for different bath velocities $V_{bath}$ = 7.0 cm/s
($\vartriangleleft$), 14.0 cm/s ($+$), 17.5 cm/s ($\lozenge$), 21.0
cm/s ($\times$). }
 \label{fig:Scaling}
\end{figure}

Consider the energy associated with the additional surface
needed to separate the rebounding jet from the bath.  The energy of
the new interfacial surface scales as an area
\begin{equation}\label{E_bouncing}
E_{bouncing} = \int {\sigma} dA \sim \sigma S^2 \, ,
\end{equation}
where $S$ is the arc-length of the roughly ellipsoidal indentation
on the bath. The surface area present before bouncing only change the coefficient, not the
scaling.  Additionally, the length scale $S$ scales linearly with
$Q$ with a slope $b$ = 3.13 s/cm$^2$, as shown in the inset in
Fig.~\ref{fig:Scaling}.

Now consider the energy of the jet plunging into the bath. The jet slows down, widens, and is
carried along by the bath; the trumpet-shaped air film can be seen
in Fig.~\ref{fig:Multiplicity}(a). We model the plunging jet simply
as a straight vertical cylinder of length $L$ with a diameter
$d_{jet}$ moving through a fluid at velocity $V_{bath}$
perpendicular to the cylinder's axis; this very rough model is used
just to obtain some indication of parameter dependencies. The Stokes drag force $F_{drag}$ exerted
on the cylinder is
\begin{equation}
F_{drag} =  \frac{4 \pi \mu V_{bath} L}{ln(7.4/Re_{bath})},
\end{equation}
where $Re_{bath} = d_{jet} \rho V_{bath} / \mu$  \cite{Lamb}. The
associated work is
\begin{equation}
E_{plunging} = \int F_{drag} dx \propto F_{drag} \ell \,
\end{equation}
where $\ell$ is the length of integration.  The length $\ell$ is
taken as the horizontal distance that the bath advects a parcel of
the jet during the time $\Delta t$ that it traverses the cylinder of
length $L$ traveling with velocity $V_{jet}$,
\begin{equation}
\ell = V_{bath} \Delta t = V_{bath} \frac{L}{V_{jet}}.
\end{equation}

The length of the cylinder $L$ is expected to increase with $Q$ and
$V_{jet}$, and decrease with $\mu$ and $V_{bath}$. In accord with the
expected functional dependence of the cylinder length, we take as an
ansatz
\begin{equation}\label{length}
L \sim a \sqrt{\frac{Q}{V_{bath}}},
\end{equation}
where $a$ = 20 (dimensionless) makes $L$ comparable to estimates of the
effective jet length, which is longer than the air sheath (typically 0.5-2
cm) since the jet penetrates farther than the air sheath.  With this
factor $a$ the energies $E_{plunging}$ and $E_{bouncing}$ are comparable,
but given the neglected coefficients and many rough approximations of our
model, the energy magnitudes are very uncertain.

Substituting $\ell$ and $L$ into the expression for the energy of plunging yields
\begin{equation} \label{E_plunging}
E_{plunging} \sim  \frac{4 \pi \mu V_{bath} a^2 Q}{ V_{jet}
ln(7.4/Re_{bath}) }.
\end{equation}
Because most of the jet's velocity is due to gravity and the data
are taken at the same height $H$, we assume $V_{jet}$ is constant.

In our model, the energy of the drag and of the new surface are
balanced at the transition between plunging and bouncing,
\begin{equation}
E_{plunging}=E_{bouncing}.
\end{equation}
Using the relations $S = bQ$, we have
\begin{equation}\label{result}
\frac{4 \pi \mu V_{bath} a^2 Q}{V_{jet} ln(7.4/Re_{bath})}  \sim \sigma
b^2 Q^2.
\end{equation}
The Reynolds number of the bath, which ranged from $10^{-2}$ to $6$,
was calculated using $d_{jet} = \sqrt{Q} \sqrt{4/(\pi V_{jet})}$.
Our measurements of the transition curves at four different
viscosity values collapse well when expressed as in (\ref{result}),
as Fig.~\ref{fig:Scaling} illustrates. The data indicate
that as $Q$ increases, $E_{plunging}$ grows faster than
$E_{bouncing}.$  However, a power law fit yields an exponent of $1.6 \pm 0.1$, where the uncertainty is the deviation of power law fits of the individual transition curves where they have
positive slope.

\section{\label{sec:Kitchen}Kitchen Experiments}

The bouncing jet phenomenon can be observed in many household fluids
such as canola oil or heavy mineral oil.  Bouncing was first
observed in our laboratory while pouring silicone oil by hand into a
dish for storage. The materials needed for observing a bouncing jet
are simple: a dish (preferably transparent like a glass pie pan, at
least 15 cm in diameter and 4 cm tall), a cup, and a small rod (e.g.
a cable tie or a chopstick).  We measured the viscosity of canola
oil and heavy mineral oil (at 22$^\circ$ C) at high shear rates  and
found both oils to be Newtonian to a good approximation: for canola
oil, $\mu$ = 65 mPa s at low shear and 4\% lower at a shear of
10$^4$ s$^{-1}$; for heavy mineral oil, $\mu$ = 180 mPa s at low
shear and 18\% lower at a shear rate of 10$^4$ s$^{-1}$.

To observe a bouncing jet, use a dish with liquid about 4 cm deep
and pour a thin stream of the liquid (0.5 to 1 cm$^3$/s) from a cup
3 to 6 cm above the surface. While pouring, move the stream in a
circular motion around the dish once about every 2 seconds at a
distance 3 to 6 cm from the center.  Watch for the jet to bounce
while varying the pouring rate, the relative horizontal velocity
between the jet and the bath, and the pouring height. To encourage
bouncing, pass the small rod through the jet intermittently. A
rotating platform (e.g. a record turntable or a Lazy Susan) can be
used to rotate the dish instead of moving the cup.  If the surface
is dirty, clean the surface by stirring the bath or scraping the
surface \cite{noncoal:rayleigh3,noncoal:dropss}. Blow air on the
surface to pop bubbles on the surface.  To achieve a very small
pouring height, pour the liquid down the rod.

With practice, a jet poured by hand can bounce stably for tens of
seconds at a time.  For more detailed instructions and for more
liquids and conditions, see \cite{thesis}. Bouncing is also easy to
observe in non-Newtonian fluids such shampoo, multigrade motor oil,
and concentrated mixtures of liquid soap and water, but the
mechanism by which they bounce may not be the same as for Newtonian
fluids.

\section{\label{sec:Discussion}Discussion}

We have observed a falling jet of a Newtonian liquid bouncing from a
moving bath of the same liquid for a wide range of viscosity (52 to
349 mPa s), jet diameter (0.05 to 0.12 cm), jet velocity at impact
(38 to 170 cm/s), and the bath's horizontal velocity (0.5 to 35
cm/s).  By initiating the jet in different ways, as many as four
stable states were observed for the same experimental conditions.

The bouncing jet is a new example of steady non-coalescence and a
new example of a fluid flow with multiple stable states. Bouncing
jets could be used as a new technique for controlling a fluid jet
and preventing or promoting the entrainment of the fluid surrounding
the jet. The phenomenon can be observed easily with a variety of
fluids at home.

The bouncing phenomenon we have studied is similar to the Kaye
effect: both are thin streams of liquid rebounding from a surface,
and both occur for similar falling heights, jet velocities, and jet
diameters. However, the liquid of the stable Kaye effect is
non-Newtonian, while we have studied bouncing for Newtownian
liquids. The Newtonian bouncing liquid jet is separated from the
bath by an air layer, likely 0.1 to 10 $\upmu$m thick (see
discussion in Section \ref{subsec:VbandQ})
~\cite{noncoal:review2,noncoal:mason,entrap:quere}. In contrast, the
stable Kaye effect is lubricated by a shear-thinned layer of liquid
about 100 $\upmu$m thick~\cite{kaye:lohse}.  The bouncing jet also
occurs for less viscous liquids than the Kaye effect.

The behavior of a bouncing jet is determined by an interplay of
viscous, inertial, surface, and gravitational forces. This causes
the relationship between any two features of the bouncing jet to be
very complicated.  We have suggested that the onset of bouncing can
be approached by comparing the energies of drag on the plunging jet
and the energy of the new surface of the bouncing jet. While the
argument collapses the transition curves, it cannot explain the
slope of the curves or their closure at small $Q$.  To understand
this transition better, future experiments should examine the
dependence on other parameters, including surface tension, density
difference, radius of the nozzle, angle of incidence of the jet, and
the viscosity and pressure of the surrounding fluid (which was air
at atmospheric pressure in our experiments).  Many of these
conditions have already been studied for the case of droplet
non-coalescence \cite{noncoal:review}. In addition, we have observed
that a falling jet impinging on a moving bath exhibits other
phenomena that warrant future study \cite{thesis}.

\begin{acknowledgments}
We thank W.D. McCormick and J.W.M. Bush for helpful discussions. We
also thank Olivier Praud and Abe Yarbrough for help with preliminary
experiments. Acknowledgment is made to the Donors of the American
Chemical Society Petroleum Research Fund for support of this
research.
\end{acknowledgments}

\bibliography{jet070625}

\end{document}